# ANALYSIS OF 3GM CALLISTO GRAVITY EXPERIMENT OF THE JUICE MISSION

Mauro Di Benedetto,[*] Paolo Cappuccio,[†] Serena Molli,[‡] Lorenzo Federici[§], and Alessandro Zavoli[**]

The ESA's JUICE mission will provide a thorough investigation of the Jupiter system and the Galilean moons during its nominal tour, comprising flybys of Europa and Callisto, and an orbital phase about Ganymede at the end of the mission. The 3GM experiment will exploit accurate Doppler and range measurements to determine the moons' orbits and gravity fields (both static and tidal) and infer their interior structure. This paper presents the attainable accuracies of the Callisto geodesy experiment and addresses the effect of different flybys mean anomaly distribution and geometry on the estimation of the tidal Love number $k_2$.

## INTRODUCTION

JUICE (JUpiter ICy moons Explorer) is the first ESA L-class mission devoted to the exploration of Jupiter and its icy moons. The current nominal launch date is in mid 2022 and the arrival at Jupiter is expected at the end of 2029, after a 7.4 years interplanetary phase and an Earth-Venus-Earth-Mars-Earth (EVEME) gravity assist sequence. After an initial Ganymede flyby a few hours before the capture maneuver (to reduce the size of this latter), JUICE will initiate its nominal Jovian tour and the mission science phase.

Ganymede is the primary scientific objective of the entire mission and will be thoroughly studied during a 9-months orbital phase at the end of the Jovian tour. Before entering in orbit around Ganymede, JUICE will perform several flybys of Europa, Ganymede and Callisto, gathering precious multidisciplinary data with its payload composed by 10 different scientific instruments. Shortly after the capture maneuver, JUICE will be put in resonant orbits with Ganymede, in the so-called energy reduction phase, to reduce further the orbital energy and start the Europa science phase. JUICE will perform only two flybys of Europa during the entire mission, being limited on the possibility of a greater number by the harsh radiation environment close to the moon's orbit. Then it will be put in resonant orbits with Callisto, using repetitive gravity assists to raise (and then reduce back to the equatorial plane) the inclination of the orbit around Jupiter, and perform observations of the Jovian atmosphere at high latitudes in the northern hemisphere. The actual duration (and the number of Callisto flybys) of this high-inclination phase depends on the particular resonance chosen and the available navigation $\Delta v$[1]. The current baseline is to reach a minimum inclination of 22° with respect to the equatorial plane of Jupiter and perform a total of 12 Callisto flybys. In this mission scenario, the total duration of the inclined phase is approximately 9-10 months and such a series of flybys will be used to fulfil the requirements related to Callisto science. The comparative study of Europa, Ganymede and Callisto aims at answering two of the main themes of the ESA Cosmic Vision 2015 – 2025 program: "What are the conditions for the planet formation and emergence of life?" and "How does the Solar System work?"[2].

[*] 3GM Experiment Manager, PhD, Dipartimento di Ingegneria Meccanica e Aerospaziale, Sapienza, Università di Roma, via Eudossiana 18, 00184, Rome, Italy
[†] PhD Student of Aeronautical and Space Engineering, Dipartimento di Ingegneria Meccanica e Aerospaziale, Sapienza, Università di Roma, via Eudossiana 18, 00184, Rome, Italy
[‡] Graduate Student of Aeronautical and Space Engineering, Dipartimento di Ingegneria Meccanica e Aerospaziale, Sapienza, Università di Roma, via Eudossiana 18, 00184, Rome, Italy
[§] PhD Student of Aeronautical and Space Engineering, Dipartimento di Ingegneria Meccanica e Aerospaziale, Sapienza, Università di Roma, via Eudossiana 18, 00184, Rome, Italy
[**] Researcher, PhD, Dipartimento di Ingegneria Meccanica e Aerospaziale, Sapienza, Università di Roma, via Eudossiana 18, 00184, Rome, Italy

3GM (Gravity and Geophysics of Jupiter and the Galilean Moon) is the radio science experiment onboard the JUICE mission, comprising two separate and independent units: a Ka band Transponder (KaT) enabling very accurate two-way coherent Doppler and range measurements (34-32.5 GHz), and an Ultra Stable Oscillator (USO). The two units pertain scientific objectives in geodesy and geophysics (KaT), and atmospheric science (USO). By means of an accurate orbit determination, the KaT will allow improving the current knowledge of the interior structure of the three moons (by measuring their gravity fields[3]) and the Jovian satellite ephemerides (by solving for the spacecraft and moons orbits). The aim of this paper is to show a covariance analysis of the Callisto gravity experiment, whose goals are the estimation of the moon gravity field and its tidal response, controlled by the Love number $k_2$, to the time varying gravitational pull from Jupiter, induced by the moon's eccentric orbit around the gas giant. The accurate estimation of the Love number $k_2$ is crucial to unambiguously detect the presence of a global subsurface ocean, one of the key mission science objectives. We will also show the experiment performances attainable considering a notional mission kernel referring to a launch in 2023 and a redesign of the last Callisto flyby in the nominal trajectory. These two latter cases are used to evaluate the 3GM performances in cases which present more favorable mean anomaly distribution and geometry.

**CALLISTO GRAVITY EXPERIMENT**

**Science goals**

Measurements carried out by the Galileo spacecraft have evidenced that Callisto has the lowest mean density and at the same time the highest axial moment of inertia $C/MR^2$ among the four Galilean satellites, indicating that the moon cannot be entirely differentiated[4]. In addition, the presence of an induced field suggests that, similarly to Europa, Callisto could host a subsurface ocean[5].

The main purpose of the Callisto gravity experiment is therefore to unambiguously detect the presence of a global liquid ocean beneath the icy crust and refine modelling of the interior structure, assessing the extent of the internal differentiation and determining whether the moon is in hydrostatic equilibrium, so removing the ambiguity in the interpretation of Galileo data. In order to do that, the 3GM experiment aims at measuring:

- the tidal Love number $k_2$ with an absolute uncertainty better than $6 \times 10^{-2}$;
- Callisto degree 3 gravity field (with an accuracy for $J_3$ better than $4 \times 10^{-8}$).

The capability to meet these requirements (and possibly to go beyond them) is determined by several aspects that all play a fundamental role. Among the most important ones there are: 3GM data accuracy, tour design (affecting flybys geometry) and operational constraints imposed by the spacecraft.

**Measurement technique**

The mass and mass distribution of Callisto can be inferred through the orbit determination of the JUICE spacecraft, which acts as a test mass, during flybys. The moon's gravity field is estimated by adjusting gravity coefficients (and other dynamical and observational parameters) to minimize the sum of squared residuals (SSR), that are defined as the differences between predicted and observed observables. The computation of predicted observables requires a reference trajectory (thus a dynamical model) and an observational model. In a real experiment the vector of residuals cannot be null because actual measurements are always affected by a certain amount of noise. In our numerical simulation, we have created synthetic observed observables by adding a realistic measurement noise based on current best knowledge of radio metric error models (see next section).

Both dynamics and measurements involve nonlinear relationships that require an iterative procedure to get the weighted least square differential corrections to be added to the first guess solution. In the classical batch filter formulation of the OD problem with given a priori information, the solution for differential correction at $k$-step yields[6]:

$$\delta \hat{\mathbf{x}}_k = \left( H_k^T W_k H_k + \bar{P}_k^{-1} \right)^{-1} \left( H_k^T W_k \delta \hat{\mathbf{y}}_k + \bar{P}_k^{-1} \delta \bar{\mathbf{x}}_k \right) \tag{1}$$

where $\delta \hat{\mathbf{x}}$ is the vector of differential corrections to be used to correct **X**, the $n$-dimensional vector of the solved-for-parameters, being $n$ the number of estimated parameters. $H$ is the $p \times n$ mapping matrix ($p$ is the number of observables) containing the partial derivatives of the observables with respect to the solved-for-parameters. $W$ is the weighting matrix while $\delta \bar{\mathbf{x}}$ and $\bar{P}_k$ represent, respectively, the a priori estimate and covariance of **x**. $\delta \mathbf{y}$ is the vector of observable residuals. If data are weighted with the inverse square of the RMS of the measurement noise, that is $w_i = 1/\sigma_i^2$, being $i = 1, \dots, p$ the $i$-th observable, the first term of the right-hand side in Eq. (1)

$$P_k = \lambda_k^{-1} = \left( H_k^T W_k H_k + \bar{P}_k^{-1} \right)^{-1} \tag{2}$$

represents the minimum $n \times n$ variance-covariance matrix of the solved-for-parameters, being the square root of diagonal terms the standard deviation error $\sigma_j$ of the $j$-th estimated parameter, with $j = 1, \dots, n$. Typically, the inclusion of a priori information allows mitigating potential numerical instabilities associated to inversion of $\lambda_k$ (called information matrix).

3GM observables will be acquired during flybys occurring over a timespan of several months. In general, inaccuracies of the dynamical model, numerical rounding errors of the orbit propagators (due to finite machine number representation), and orbit manoeuvres (affecting the dynamical coherence of the JUICE trajectory) make virtually impossible to fit radio metric data in one single batch. This problem is overcome by using a multiarc approach, in which the JUICE trajectory is split into many shorter arcs, and the set of estimated parameters is divided into two categories: global and local ones[7]. Global parameters are those common to all arcs (e.g. gravity coefficients) while local ones are those specific to any single flyby (e.g. the spacecraft initial state).

From the mathematical point of view, this approach is implemented through a proper population of the mapping matrix $H$. Considering for simplicity only two arcs, we have the two solved-for-parameters vectors $\mathbf{Z_1} = [\mathbf{X_1} \quad \mathbf{X_g}]$ and $\mathbf{Z_2} = [\mathbf{X_2} \quad \mathbf{X_g}]$, where $\mathbf{X_1}$ and $\mathbf{X_2}$ are locally estimated parameters and $\mathbf{X_g}$ is the vector of global parameters. The two vectors can be merged into $\mathbf{Z} = [\mathbf{X_1} \quad \mathbf{X_2} \quad \mathbf{X_g}]$ and the $H$ matrix can be written as:

$$H = \begin{bmatrix} \dfrac{\partial y_1}{\partial x_1} & 0 & \dfrac{\partial y_1}{\partial x} \\ 0 & \dfrac{\partial y_2}{\partial x_2} & \dfrac{\partial y_2}{\partial x} \end{bmatrix} \qquad (3)$$

where $\mathbf{y}_1$ and $\mathbf{y}_2$ are data from the two arcs. The $H$ matrix reflects the fact that partial derivatives from arc 1 are null with respect to local parameters of arc 2 (and vice versa). However, the estimation covariance matrix $P_k$ is a fully populated one, with diagonal terms still representing the $j$-th parameter estimated variance $\sigma_j^2$, with $j = 1, \dots, n$.

**JUICE dynamical model**

The gravity field outside a nearly regular shaped body of mass $M$ and reference radius $a_e$ can be derived from the spherical harmonic expansion of the gravity potential $U$ in the body fixed-reference frame:

$$U(r, \lambda, \phi) = \frac{GM}{r} + \frac{GM}{r} \sum_{l=2}^{\infty} \sum_{m=0}^{l} \left(\frac{a_E}{r}\right)^l \bar{P}_{lm}(\sin \phi) \times [\bar{C}_{lm} \cos m\lambda + \bar{S}_{lm} \sin m\lambda] \qquad (4)$$

where $(\bar{C}_{lm}; \bar{S}_{lm})$ are the normalized coefficients of degree $l$ and order $m$, $G$ is the gravitational constant, $\bar{P}_{lm}$ are the fully normalized associated Legendre polynomials, while $\phi$, $\lambda$, and $r$ are, respectively, latitude, longitude and radial distance. Terms with $l = 1$ are null if one assumes the body fixed reference frame coincident with the body center of mass. Going into higher degree terms provides a finer description of gravity features on a smaller regional scale. Since the acceleration induced by the $l$-th term is $a_l \propto 1/r^{l+1}$, de facto, truncation of Eq. (4) occurs depending on 3GM measurement accuracy, the orbit geometry and the amount of available data, all affecting the parameters observability. In practice, in order to have a reliable and robust estimation of higher terms, it is always convenient to include additional terms because of the correlation among gravity coefficients. For Callisto we have run numerical simulations including till a full 8x8 gravity field, where nominal values for quadrupole coefficients are those estimated by the Galileo spacecraft while for higher order terms we have used the empirical Kaula's rule:

$$\bar{C}_l^2 = \frac{A_k 10^{-10}}{l^4} \qquad (5)$$

$$\bar{C}_l^2 = \frac{1}{2l+1} \sum_m (\bar{C}_{lm}^2 + \bar{S}_{lm}^2) \qquad (6)$$

which provides the amplitude of gravity coefficient spectrum. Eq. (5) is known to predict quite well the high order coefficients of rocky bodies, but its applicability to icy satellites is questionable. Actually, for a pure covariance analysis the attainable accuracy on the estimation of gravity coefficients is independent from their nominal value, so we may have chosen null values, also considering that the dynamical effect on the JUICE trajectory is limited on our short arc strategy. However, we have assumed a weak gravity field ($A_k = 1$) to have an indication of the minimum gravity field we may estimate with sufficient relative uncertainty.

The JUICE gravity model includes also point mass accelerations of the Sun, all solar system planets and of the four Galilean satellites. For the position of all bodies included in the JUICE gravity model we have used the DE430 planetary ephemerides model and NOE-5-2017-GAL for the Jovian satellites ephemerides.

The orbit of Callisto is eccentric ($e_C = 0.0074$) and the Jupiter's gravitational attraction varies by roughly 3% between periapsis and apoapsis. By this, Callisto experiences a time-varying tidal distortion along its orbit and its gravity field is therefore not static. The periodic variations of the moons' quadrupole field in response to Jupiter tidal potential $U_2^J$ are controlled by the Love number $k_2$:

$$\Delta U_2 = k_2 \left(\frac{a_e}{r}\right)^3 U_2^J \qquad (7)$$

Typically, a certain number of non-gravitational accelerations can perturb the orbit of a spacecraft. The multi arc approach we have used for the Callisto gravity experiment allows us to neglect any spacecraft manoeuvres, which are typically carried out far from the closest approach, so outside 3GM measurement windows. Due to the size of JUICE solar arrays (> 70 m$^2$), the perturbative acceleration induced by the solar radiation pressure $a_{SRP}$ is at level of ~ 10$^{-12}$ km/s$^2$, potentially causing a displacement of the spacecraft by a few meters over one day, much larger than KaT ranging accuracy, which is at level of some centimeters (see next section). By this, we have included the solar radiation pressure in the JUICE dynamical model and thermo-optical coefficients (specular and diffuse reflectivity) in the list of estimated parameters.

**3GM data and noise model**

The core of the 3GM gravity experiment is a state-of-the-art Ka band Transponder manufactured by Thales Alenia Space - Italy (TASI), a nearly recurrent unit from MORE's KaT onboard the ESA/BepiColombo mission to Mercury[*]. The KaT enables a very stable two-way coherent Doppler link (34-32.5 GHz), having an intrinsic fractional frequency stability better than 10$^{-15}$ at 1000 s integration time. Even better performances, at level of ~ 5×10$^{-16}$ at 1000 s integration time, could be expected relying on ground tests of the MORE's KaT flight unit and preliminary measurements already carried out on 3GM's KaT. The unit also implements a ranging channel based on Pseudo Noise (PN) codes at 24 Mcps, enabling 20 cm accuracy (two-way).

Typically, such instruments exhibit a white phase noise behavior ($S(y) \propto f^2$ where $y = \Delta f/f$) in the 3GM Doppler band of interest, e.g. [10$^{-4}$÷10$^0$] Hz, traducing into an Allan deviation $\sigma_y \propto \tau^{-1}$. Potentially, this would correspond to 2.5 μm/s measurement accuracy at 60 s integration time. However, in very precise Doppler tracking experiments the measurement accuracy is limited by other noise sources, mainly propagation noise (due to interplanetary plasma and Earth troposphere) and mechanical disturbances generated both on ground and onboard the spacecraft. Dispersive noise due to solar plasma is effectively suppressed in a wide range of Sun-Earth-probe (SEP) angles already by using the Ka band. At very low SEP angles, the simultaneous use of the 3GM Ka/Ka link and the two additional X/X (7.2-8.4 GHz) and X/Ka (7.2-32.5 GHz) coherent links provided by the onboard Deep Space Transponder (DST) allows a nearly complete calibration of the dispersive noise, making JUICE virtually immune to dispersive noises. On the opposite, calibration of the noisy effect due to the Earth troposphere, which is non-dispersive in the microwave region, requires the use of advanced water vapour radiometers installed close to the large ground antenna dish. The noise contribution from the already operational ESA DSA2 ground station in on average $\sigma_y \sim 1.6 \times 10^{-14}$ at 60 s integration time[8], but the new Malargüe DSA3 that will support Ka band uplink during 3GM experiment is expected to perform much better. The 3GM Doppler measurement accuracy will be therefore able to approach the virtual limit of the achievable performances of tracking systems in the microwave region, close to 10 µm/s accuracy at 60 s integration time[6]. In the numerical simulations of Callisto gravity experiment we have considered a bit conservative value of 12 µm/s at 60 s integration time for range-rate measurement noise and 20 cm (two-way) for ranging.

---

[*] MORE is the acronym of the Mercury Orbiter Radio science Experiment onboard the ESA BepiColombo mission.

During flybys, JUICE nadir pointing constraints settled by other instruments impose the use of a steerable 50 cm Medium Gain Antenna (MGA). Rotations of the spacecraft can excite the propellant in the two spacecraft tanks, inducing a potential disruptive noisy effect on 3GM observables (mainly on range-rate). By this, JUICE will host onboard a High Accuracy Accelerometer (HAA) for the calibration of non-gravitational perturbations due to sloshing. The HAA is a nearly recurrent unit from the Italian Spring Accelerometer (ISA) onboard BepiColombo. The HAA measurements can be modeled as:

$$a_{HAA} = \lambda a_{NGA} + b_0 + \epsilon \tag{8}$$

where $a_{NGA}$ are actual non-gravitational accelerations and $a_{HAA}$ are the HAA readings. These two quantities differ because of a measurement scale factor $\lambda$ (ideally equal to 1 but in practice never so), a bias error $b_0$ and the unit intrinsic noise level $\epsilon$. The ISA unit is designed to have $\epsilon$ at level of $10^{-9} \text{m/s}^2\sqrt{\text{Hz}}$ in the $3\times10^{-5}$ to $10^{-1}$ Hz frequency band[9], and the JUICE's HAA is expected to perform similarly. By this, the noise from the propellant sloshing is assumed to be entirely calibrated by the HAA down to a level compatible with the expected 3GM measurement accuracy. The knowledge of the scale factor and the bias cannot rely on ground measurements and requires in-flight calibration. These two quantities are therefore estimated during the OD procedure. Since the HAA has three different sensing elements in three orthogonal directions, this traduces into adding a total of six parameters: $\lambda_X, \lambda_Y, \lambda_Z, b_{0X}, b_{0Y}, b_{0Z}$. In addition to that, any time the HAA is switched on and off, these quantities may vary due to internal thermo hysteresis of the unit and require a new calibration. This is a serious complication for 3GM because the HAA calibration parameters must be treated as local parameters, significantly increasing the size of the estimation variance-covariance matrix. In the following sections we will show results of our numerical simulations under different assumption on the use of the HAA.

**Simulation setup**

Ideally, gravity measurements during flybys are carried out by performing at least three spacecraft Doppler tracking passes, one at closest approach and two at 'wings' (to better constrain the inbound and outbound arc segments) over a dynamically quiet arc (a crucial pre-requisite for any radio science observation). Since ESA/ESTRACK ground complexes have currently only one Deep Space Antenna (DSA) equipped with Ka band uplink capabilities (DSA3 in Malargüe), establishing a coherent two-way Ka/Ka link at wings would require a clean arc for roughly two days. However, at least for some Callisto flybys, wheel off-loading desaturation manoeuvres due to momentum torque disturbances may prevent the possibility to have two-days clean arcs. Preliminary analyses have shown that, conservatively, only a shorter quiet period (~ 30 hrs) can be ensured. By this, in this work we have considered this arc length, with initial spacecraft state vectors taken 15 hrs before any closest approach and extracted from two notional mission kernels made available by the JUICE project: case 141a (baseline) and case 230la (backup). The backup trajectory refers to a launch date postponed to August 2023 and a longer cruise duration (9 years instead of 7.4). Figure 1 and 2 show the mean anomaly distribution for such two cases, while Figure 3 shows the different ground tracks for the two different trajectories.

The DSA3 can be in visibility only for 6-8 hrs per day (a typical tracking pass duration), so only an X/X or possibly a X/Ka link can be established during the rest of the flyby. Far from superior solar conjunctions, the solar interplanetary plasma (and to a less extent the Earth ionosphere) is not a leading noise source and performances of X and Ka links become comparable. By this, in our simulations we have considered a nearly continuous tracking from ground over the entire arc and a constant link stability, with an Allan deviation $\sigma_y \sim 10^{-14}$ at 1000 s integration time. Assuming a white gaussian noise in the $10^{-4} \div 10^0$ Hz frequency band, as already experienced in many precise spacecraft Doppler tracking experiments, $\sigma_y \sim \tau^{-0.5}$ which corresponds to a range-rate accuracy of 12 μm/s at 60 s integration time. We have also investigated solutions obtained with the inclusion of range data (one range data point each 300 s), having an accuracy of 20 cm two-way. We have discarded from the data set any range and range-rate data observable having an elevation lower than 15° because of the increasing noisy effect of the Earth troposphere.

The set of estimated parameters included in the OD filter is reported in Table 3 along with their a priori uncertainties. The a priori knowledge of HAA calibration parameters is derived from measurements carried out by the instrument manufacturer (TAS-I). As already pointed out, if the HAA is powered continuously over the entire duration of the Callisto phase such two quantities can be regarded as global parameters, while any time the unit is switched off they shall be newly estimated as if we were dealing with a 'new' unit. We have also included thermo-optic coefficients (diffuse and specular) of the JUICE solar arrays. The nominal value may be not representative of actual surface finishing, but this is not relevant for a pure covariance analysis. Since range measurements are more prone than Doppler to error biases, we have also included in our solution one different range bias parameter at any tracking pass.

**Table 1. List of Callisto flybys (case 141a)**

| Flyby | Date | Mean anomaly | Altitude [km] |
|---|---|---|---|
| C1 | 13/10/30 | -57° | 412 |
| C2 | 13/12/30 | -174° | 200 |
| C3 | 30/12/30 | -175° | 200 |
| C4 | 15/01/31 | -177° | 200 |
| C5 | 01/02/31 | -174° | 200 |
| C6 | 25/04/31 | -177° | 200 |
| C7 | 12/05/31 | -176° | 200 |
| C8 | 29/05/31 | -175° | 1020 |
| C9 | 14/06/31 | -178° | 1547 |
| C10 | 01/07/31 | -176° | 200 |
| C11 | 27/09/31 | -87° | 200 |
| C12 | 25/11/31 | 107° | 3414 |

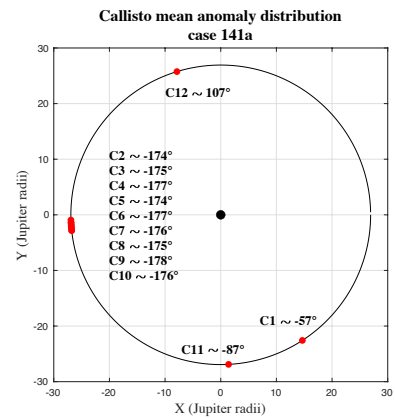

**Figure 1. Callisto mean anomaly distribution (case 141a)**

**Table 2. List of Callisto flybys (case 230la)**

| Flyby | Date | Mean anomaly | Altitude [km] |
|---|---|---|---|
| C1 | 28/07/33 | -39° | 300 |
| C2 | 13/08/33 | -38° | 200 |
| C3 | 30/08/33 | -37° | 200 |
| C4 | 16/09/33 | -41° | 200 |
| C5 | 02/01/34 | 137° | 550 |
| C6 | 26/03/34 | 135° | 850 |
| C7 | 12/04/34 | 137° | 320 |
| C8 | 29/04/34 | 137° | 300 |
| C9 | 16/05/34 | 136° | 300 |
| C10 | 01/06/34 | 134° | 300 |
| C11 | 28/06/34 | -5° | 1550 |
| C12 | 15/07/34 | -3° | 1790 |
| C13 | 22/08/34 | 91° | 760 |
| C14 | 26/10/34 | 61° | 2000 |
| C15 | 12/09/34 | 62° | 2900 |
| C16 | 17/02/35 | 5° | 1560 |
| C17 | 04/05/35 | -160° | 2600 |

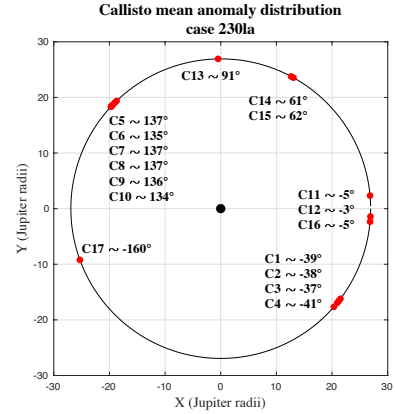

**Figure 2. Callisto mean anomaly distribution (case 230la)**

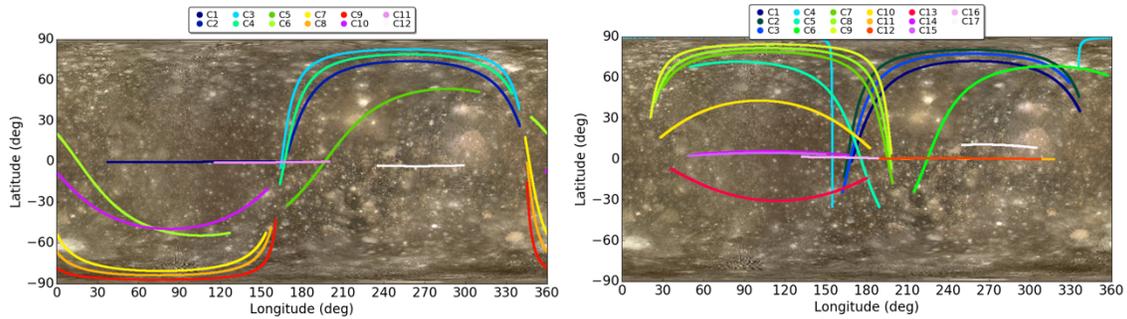

**Figure 3. Ground tracks of Callisto flybys for case 141a (left panel) and 230la (right panel).**

**Table 3. List of estimated parameters in the OD solutions and their a priori uncertainty. The HAA calibration parameters (the scale factor and the bias) can be treated as global parameters only if the unit is powered continuously over the entire experiment. If the unit is switched off the two parameters shall be newly estimated, and they shall be treated as local parameters**

| Physical quantity | Parameter | Type | Nominal Value | A priori uncertainty |
|---|---|---|---|---|
| JUICE state | Position | local | - | unconstrained |
| | Velocity | | - | unconstrained |
| Callisto state | Position | global | - | unconstrained |
| | Velocity | | - | unconstrained |
| Gravity field | Gravity coefficients | global | $J_2$ and $C_{22}$ from Galileo. Higher term from Kaula's rule ($A_k = 1$) | unconstrained |
| Gravity Tide | $k_2$ | global | 0.5 | 1 |
| HAA cal. par. | Scale factor | global/local | 1 | 0.01 |
| | Bias | global/local | 0 | $10^{-11}$ km/s$^2$ |
| Thermo-optic coefficients | Specular | global | 0.12 | 0.5 |
| | Diffuse | global | 0.12 | 0.5 |
| Range bias | | local (1 per tracking pass) | 0 | 1.2 m |

## NUMERICAL RESULTS

The gravity field of Ganymede is mostly unknown, and we still do not know the minimum degree and order required to flatten the residuals when we will get real data. By this we have run different cases going from a full 5x5 to a full 8x8 gravity field. The Figure 4 shows the attainable estimation accuracy ($1\sigma$ the red curve and $3\sigma$ the red curve with markers) compared to the simulated gravity field (blue curve) for the two 5x5 and 8x8 cases (intermediate solutions have been not reported here) referred to the 141a trajectory. The Figure 5 shows the equivalent plots obtained by considering the 230la trajectory.

In what the two solutions differ significantly is the attainable accuracy in the estimation of the Love number $k_2$. For trajectory 141a the $1\sigma$ formal uncertainty varies from ~ 0.08 (5x5 gravity field and HAA calibration parameters treated as purely global quantities) to $\sigma_{k_2}$ > 0.24, when a full 8x8 gravity field is used and HAA calibration parameters are treated as local ones (corresponding to switching off the HAA after any Callisto flyby). In the most unfavourable scenario, the estimation uncertainty becomes too large to unambiguously detect the presence of a subsurface ocean. We have also investigated an intermediate, and actually more realistic, case, in which the HAA is powered continuously during the first 5 flybys (1 to 5), then switched off and newly powered continuously for the second batch of 5 flybys (6 to 10). Finally, the unit is powered again for the last two flybys (11 &12). The rationale behind this grouping is the flybys' epoch (see Table 1). In this case $\sigma_{k_2}$ decreases to ~ 0.19 (roughly a 30% improvement). When considering the backup trajectory 230la the attainable accuracy varies from ~ 0.05 (8x8 gravity field and the HAA continuously powered) to ~ 0.06 (8x8 gravity field and HAA calibration parameters treated as local ones). Cases for 5x5 gravity field clearly provide even better results (at level of $\sigma_{k_2} < 0.04$).

The results of our numerical simulations show that for the Callisto gravity experiment the backup trajectory 230la is much more favourable and provide higher experiment margins. This was actually expected due to the large number of flybys and the much better mean anomaly distribution which provides a better sampling of the moon's gravity field.

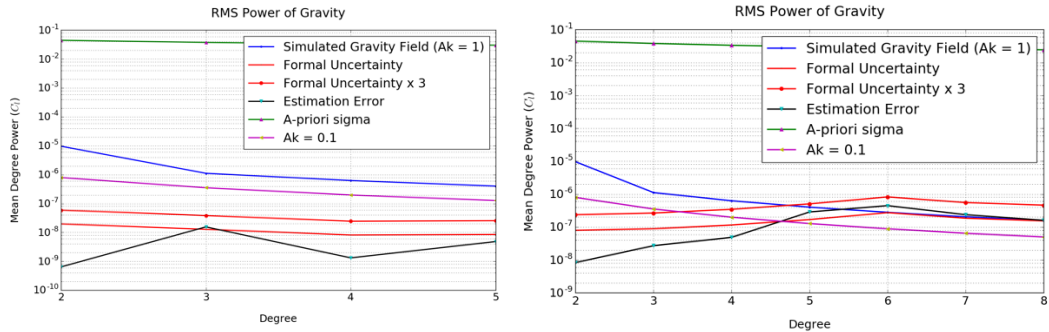

**Figure 4. Estimation accuracy (red curve $1\sigma$) attainable when considering a full 5x5 gravity field (left plot) and a full 8x8 one (right plot). Both cases refer to trajectory 141a**

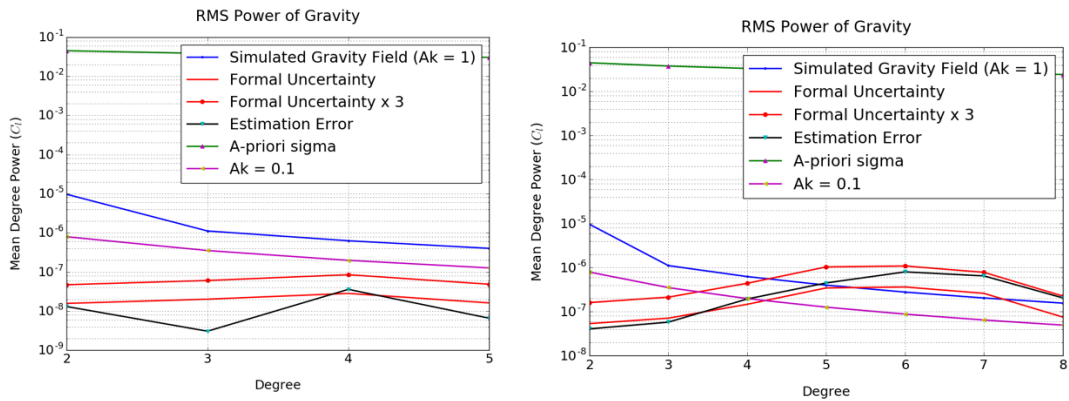

**Figure 5. Estimation accuracy (red curve $1\sigma$) attainable when considering a full 5x5 gravity field (left plot) and a full 8x8 one (right plot). Both cases refer to trajectory 230la**

**Endgame redesign**

The results shown in the previous section motivated the investigation of a redesign of the JUICE endgame trajectory in the 141a case, from after C11 flyby to GOI, so as to examine the attainable experiment performances in case the altitude of C12 flyby is constrained at $h \leq 1500$ km (while being ~ 3400 km in the nominal trajectory, see Table 1).

The trajectory redesign has been carried out assuming a patched conics model for the spacecraft dynamics and an impulsive thrust model. An MGA-1DSM (Multiple Gravity Assist – One Deep Space Maneuver) formulation is considered[10]: the trajectory is subdivided into a series of body-to-body legs, each one starting with a flyby and composed of two ballistic arcs, a *propagation arc* and a *Lambert arc*, joined by an impulsive maneuver. Spacecraft position and velocity after the C11 flyby, matching those on the 141a nominal trajectory, are used as initial conditions. This parameterization requires four parameters per leg, that are, respectively, the flyby radius, the flyby plane orientation, the leg flight-time, and the epoch of the leg flight-time at which the DSM occurs. By properly selecting the bounds of flyby and DSM epochs, it is possible to easily enforce a given resonance of the probe with a Jovian moon. Larger bounds are instead needed for transfers involving different moons, that, unfortunately, calls for a large number of local optima.

The solution of the resulting global optimization problem is carried out by means of an in-house code, named *EOS*[11], that implements a multi-population, self-adaptive, ε-constrained differential evolution algorithm, with a synchronous island-model for parallel computation. This code has been successfully applied in a broad range of space trajectory optimization problems, such as geocentric transfers, interplanetary trajectories, and rocket ascent trajectories.

**Table 4. Nominal trajectory: summary of the tour**

| Event | Date [MJD] | $\Delta t$ since last event [days] | $\Delta V$ [m/s] | C/A [km] | $v_\infty$ [km/s] | n:m post flyby |
|---|---|---|---|---|---|---|
| start | 63136.40 | 0.00 | 0.00 | / | 2.25 | 4:3 |
| C12 | 63195.64 | 59.24 | 1.02 | 3971.37 | 2.25 | / |
| G10 | 63244.73 | 49.09 | 79.42 | 19811.32 | 1.69 | 7:4 |
| G11 | 63295.30 | 50.57 | 79.81 | 33819.05 | 1.69 | 5:3 |
| G12 | 63330.34 | 35.05 | 80.12 | 53102.36 | 1.68 | 8:5 |
| G13 | 63387.95 | 57.60 | 117.94 | 6940.31 | 1.40 | 3:2 |
| G14 | 63409.08 | 21.13 | 141.82 | 15450.22 | 1.22 | 7:5 |
| G15 | 63458.88 | 49.80 | 164.53 | 33443.77 | 1.06 | 4:3 |
| GOI | 63487.13 | 28.25 | 183.59 | / | 0.91 | / |

This approach provides a reasonable estimate of the additional $\Delta V$ of the alternative mission profiles with respect to the nominal trajectory, that, according to this simplified model, requires a cumulative $\Delta V_{nom}$ = 183.59 m/s while the total endgame time is $\Delta t_{nom}$ = 350.73 d (see Table 4).

**Table 5. Additional $\Delta V$ and time in the four alternative endgames proposed from C12 to GOI**

| | $\Delta V_{add}$ [m/s] | $\Delta t_{add}$ [days] |
|---|---|---|
| Endgame 1 | 103.65 | 22.07 |
| Endgame 2 | -64.63 | 37.02 |
| Endgame 3 | -11.08 | -6.38 |
| Endgame 4 | 18.98 | -71.95 |

Table 5 summarizes the additional $\Delta V_{add}$ and the time needed from C11 to the GOI in our four alternative options of the JUICE endgame. The first option considers the same nominal flyby sequence but the time of flight between C12 and G10 is left free. The second option has been obtained by adding one additional Callisto flyby between C12 and G10 and considering the same sequence of resonant orbits with Ganymede after G10. The third option is an evolution of our second endgame, in which the 7:4 resonant orbits with Ganymede, after the inclusion of the additional Callisto flyby, have been removed. The last trajectory has been obtained by considering a Callisto-Ganymede-Callisto-Ganymede flyby sequence.

As already pointed out, we have not carried out further analyses about all likely implications of the four proposed solutions (e.g. radiation budget and scientific constrains from other instruments during the endgame). Our main purpose has been only to assess what is achievable in the 141a trajectory if a more favorable flyby geometry is possible.

It is however already noteworthy that our preliminary endgame 3 foresees a reduction of the navigation $\Delta V$ and an anticipated GOI. We will further analyze this solution, shown in the Figure 6 below, taking into account a more accurate spacecraft dynamical model.

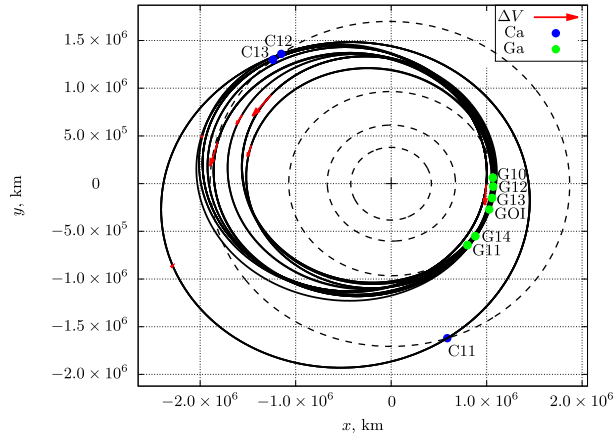

**Figure 6. Juice Endgame 3 trajectory, view from z-axis of J2000 reference frame.**

On average, the reduction from $h_{C12} \sim 3400$ km to $h_{C12} \sim 1500$ km in the 141a baseline trajectory provides an improvement in the absolute estimation uncertainty of the Love number $k_2$ of about $\Delta \sigma_{k_2} \sim 0.2$-$0.3$. This is still not enough to unambiguously detect the presence of a subsurface ocean if the HAA is switched off after any flyby and a full 8x8 gravity field is required to flatten the residuals, because $\sigma_{k_2} \sim 0.2$. However, in the more favorable scenario considering the HAA continuously powered during the three batches of flybys (1-5, 5-10 and 11-12) and a full 5x5 gravity field $\sigma_{k_2} \sim 0.06$, which actually meets the science requirement.

**CONCLUSIONS**

In this paper we have shown the attainable performance of the 3GM gravity experiment onboard the JUICE mission when the current nominal and a backup trajectory are considered. Our results show that the backup trajectory, which has a much more favourable mean anomaly distribution and geometry, provides better estimates and more margin in the estimation of the tidal Love number $k_2$, a crucial parameter to unambiguously detect the presence of a subsurface ocean. We have also shown the attainable performances in case the altitude of C12 would be reduced, providing a preliminary budget for the mission endgame re-design which foresees a reduction of the navigation $\Delta V$.

**ACKNOWLEDGMENTS**

The work described in this paper has been partially funded by the Italian Space Agency (ASI) under the contract no. 2018-25-HH.0 "Scientific activities for JUICE phase C/D".

**REFERENCES**

[1]JUI-ESOC-MOC-RP-001 – Jupiter Icy moons Explorer Consolidated Report on Mission Analysis (CReMA), issue 4.1, 18/09/2018

[2]O. Grasset, et al. "JUpiter ICy moons Explorer (JUICE): An ESA mission to orbit Ganymede and to characterise the Jupiter system". Planetary and Space Science Vol. 78, 1-21 (2013). http://dx.doi.org/10.1016/j.pss.2012.12.002

[3]P. Cappuccio, et al. "Ganymede's Gravity, Tides and Rotational State from JUICE's 3GM Experiment Simulation." Planetary and Space Science, Vol. 187, (2020). https://doi.org/10.1016/j.pss.2020.104902

[4]J. D. Anderson, et al. "Shape, mean radius, gravity field, and interior structure of Callisto". Icarus 153, 157-161 (2001a). doi:10.1006/icar.2001.6664

[5]C. Zimmer, et al. "Subsurface Oceans on Europa and Callisto: Constraints from Galileo Magnetometer Observations". Icarus Vol. 147 Iss. 2, 329-347 (2000). doi:10.1006/icar.2000.6456

[6]B. Tapley, et al., *Statistical Orbit Determination*, Academic Press, Burlington, 2004

[7]A. Milani, and G. Gronchi, Theory of Orbit Determination, Cambridge University Press (2010)

[8]L. Iess, et al. "Astra: Interdisciplinary study on enhancement of the end-to-end accuracy for spacecraft tracking techniques". Acta Astronautica 94 (2014) 699–707. http://dx.doi.org/10.1016/j.actaastro.2013.06.011

[9]V. Iafolla, et al. "Italian Spring Accelerometer (ISA): A fundamental support to BepiColombo Radio Science Experiments". Planetary and Space Science 58 (2010) 300–308. doi:10.1016/j.pss.2009.04.005

[10]L. Federici, et al. "Preliminary Capture Trajectory Design for Europa Tomography Probe," International Journal of Aerospace Engineering, vol. 2018, Article ID 6890173, 12 pages, 2018. https://doi.org/10.1155/2018/6890173.

[11]L. Federici, et al. "EOS: a parallel, self-adaptive, multi-population evolutionary algorithm for constrained global optimization,"2020 IEEE Congress on Evolutionary Computation (IEEE World Congress on Computational Intelligence), IEEE, 2020